\documentclass[proceedings, preprint]{rmaa}

\usepackage{paralist}

\usepackage{psfrag,color}

\SetYear{2016}
\SetConfTitle{ADELA}

  \title{Near-IR trigonometric parallaxes of nearby stars in the Galactic plane using the VVV \thanks{Based on observations taken within the ESO VISTA Public Survey VVV, Programme ID 179.B-2002}}
  
  \author{ J. C. Beam\'{i}n\altaffilmark{1,2},  R. A. Mendez\altaffilmark{3,2}, R. L. Smart\altaffilmark{4}, R. Jara\altaffilmark{3}, R. Kurtev\altaffilmark{1,2}, 
  M. Gromadzki\altaffilmark{5},  V. Villanueva\altaffilmark{1}, D. Minniti\altaffilmark{6,2,7}, L.C. Smith\altaffilmark{8}, P. W. Lucas\altaffilmark{8}}

  \altaffiltext{1}{Instituto de F\'isica y Astronom\'ia, Facultad de Ciencias, 
  Universidad de Valpara\'iso, Ave. Gran Breta\~na 1111, Playa Ancha, Valpara\'iso, Chile}
  \altaffiltext{2}{Millennium Institute of Astrophysics, Chile.}
  \altaffiltext{3}{Universidad de Chile, Departamento de Astronom\'ia, Casilla 36-D, Santiago, Chile.}
  \altaffiltext{4}{INAF-Osservatorio Astrofisico di Torino, Strada Osservatorio 20, 10025 Pino Torinese, TO, Italy.}
  \altaffiltext{5}{Warsaw University Astronomical Observatory, Al. Ujazdowskie 4, 00-478 Warszawa, Poland}
  \altaffiltext{6}{Facultad de Ciencias Exactas, Universidad Andres Bello, Fernandez Concha 700, Las Condes, Santiago, Chile}
  \altaffiltext{7}{Vatican Observatory, V00120 Vatican City State, Italy}
  \altaffiltext{8}{Centre for Astrophysics Research, Science and Technology Research Institute, University of Hertfordshire, Hatfield AL10 9AB, UK}
  
\shortauthor{Beam\'in, Mendez, Smart, et al.}
\shorttitle{Parallaxes from VVV survey}

 \listofauthors{J. C. Beam\'in \& R. A. Mendez \& R. L. Smart \& R. Jara \& R. Kurtev \& M. Gromadzki \& V. Villanueva \& D. Minniti \& L.C. Smith \& P. W. Lucas }
  \indexauthor{Beam\'in, J. C.}
  \indexauthor{Mendez, R. A.}
  \indexauthor{Smart, R. L.}
  \indexauthor{Jara, R.}
  \indexauthor{Kurtev, R.}
  \indexauthor{Gromadzki, M.}
  \indexauthor{Villanueva, V.}
  \indexauthor{Minniti, D.}
  \indexauthor{Smith, L. C.}
  \indexauthor{Lucas, P. W.}
\abstract{
We use the multi-epoch K$\rm _S$ band observations, covering a $\sim$5 years baseline to obtain milli and sub-milli arcsec precision astrometry for a sample of
eighteen previously known high proper motion sources, including precise parallaxes for these sources for the first time. In this study we show
the capability of the VVV project to measure high precision trigonometric parallaxes for very low mass stars (VLMS) up to distances of $\sim$250\,pc
reaching farther than most other ground based surveys or space missions for these types of stars. Additionally, we used spectral energy distribution to
search for evidence of unresolved binary systems and cool sub-dwarfs. We detected five systems that are most likely VLMS belonging to the Galactic
halo based on their tangential velocities, and four objects within 60 pc that are likely members of the thick disk. A more comprehensive study of high
proper motion sources and parallaxes of VLMS and brown dwarfs with the VVV is ongoing , including thousands of newly discovered objects.}

\resumen{}

\addkeyword{Methods: Astrometry}
\addkeyword{Stars: low mass}
\addkeyword{Stars: Brown dwarfs}
\addkeyword{Stars: subdwarfs}
\addkeyword{Methods: Proper motions}

\begin{document}
\maketitle

\section{Introduction}
\label{sec:intro}
The VVV survey is a multi-epoch near infrared ESO public
survey using the VISTA 4.2 m telescope, observing over
560 sq. degrees towards the most crowded regions of the
sky, the galactic bulge and disk.
VVV mapped these areas in five NIR bands (ZY JHKs) and
had a variability campaign lasting 5-6 years in the Ks band,
obtaining between $\sim$70 and $\sim$150 epochs per pointing.
Additionally, the small pixel size of 0.34'' (and a typical
FWHM of 1'' in the Ks band) provide a good resolution
and make the VVV survey ideal to find new moving
sources in the most crowded areas in the sky. 

Taking advantage of all these characteristics we performed
a search for high proper motions sources and we found
thousands of them \citep{Kurtev2017,Smith2016}.

We were able to obtain spectroscopic follow up
observations for several tens of sources and the results
were discussed in \citealt{Beamin2015} and  \citealt{Gromadzki2016}, 
most of them K-M dwarfs. Additionally a new sample of $\sim$50 ultra cool dwarfs (UCDs) with full atrometric characterization and classification with low-mid Near IR resoluion spectra from the VVV survey is being analyzed and made publicly available soon (Beamin et. al. in prep). 

We selected sources from previous studies of high proper
motion (HPM), lying within the VVV area that were not saturated on the images.
We obtained ten sources from the studies of \citet{Lepine2005}
and \citet{Lepine2008}, and seven co-moving companions discovered
by \citet{Ivanov2013}, and VVV BD001 \citet{Beamin2013}

Among our discoveries two of them are particularly
interesting an unusually blue L dwarf at 18.6 pc (revised
distance, this work) and a benchmark early L dwarf
companion to the $\sim$400 Myr star $\beta$ Circini \citep{Smith2015}.

In this work we focus in obtaining precise parallax
measurements for a sample of cool stars within 250 pc
from the sun. Some of the stars we selected will also be
observed by Gaia, allowing us to crosscheck the
astrometric accuracy of the VVV survey. 

\section{Methds}
\label{sec:methods}
\subsection{Target Selection}
Target selection:
We selected sources from previous studies of high proper
motion (HPM), particularly, we selected all the sources in
the VVV area that were not saturated in the VVV images.
We obtained ten sources from the studies of \citealt{Lepine2005} and \citealt{Lepine2008}, seven co-moving companions discovered by \citealt{Ivanov2013}, and VVV BD001 from \citealt{Beamin2013}. 
\subsection{Astrometric procedure}
For these eighteen sources we selected reference stars in
a circle of radius 1' around the target which, given the high
stellar density in these fields, provides an adequate
number of reference stars (above a hundred sources).
These were selected among the highest S/N objects in the
field of view, satisfying the condition that they appear in at
least 80$\%$ of the frames, and do not exhibit large proper motions.

After all epochs were on the same astrometric reference
system, we fit proper motion, parallax and a shift per
coordinate. More details on the procedure can be found in 
\citet{Smart2003} and \citet{Beamin2015}.

\section{Results}
\label{sec:results}
\begin{figure}[!t]
  \includegraphics[width=\columnwidth]{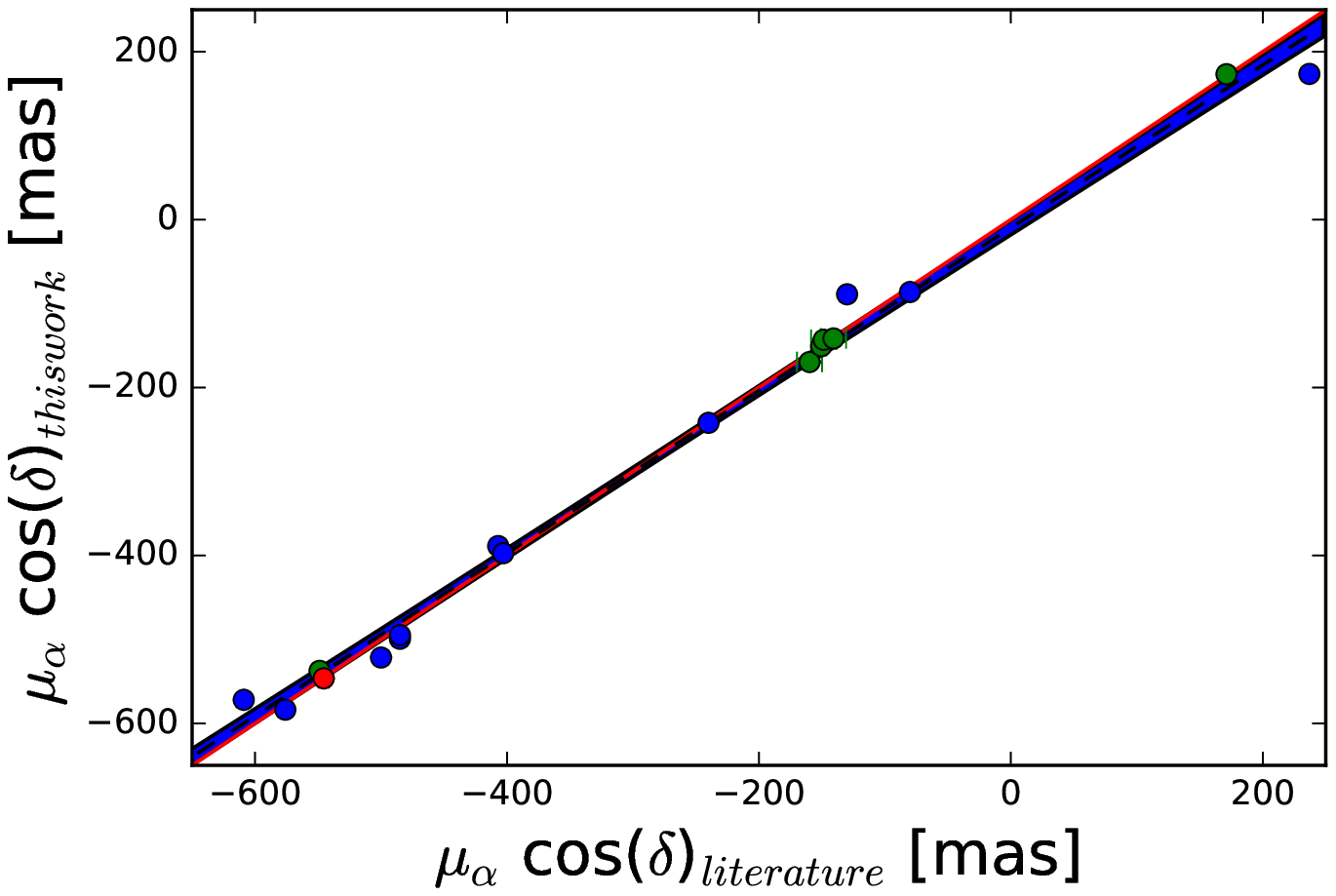}
   \includegraphics[width=\columnwidth]{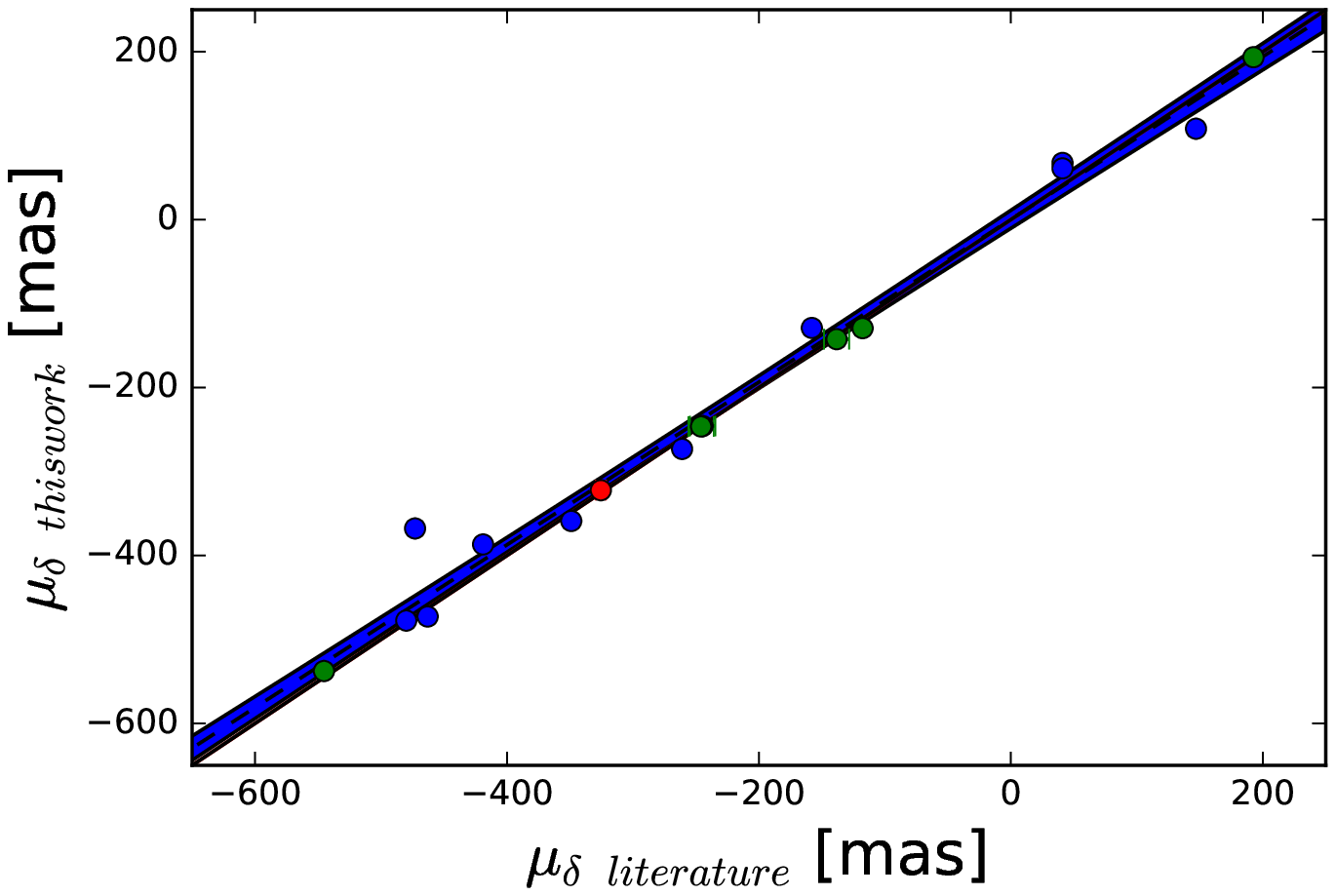}
  \caption{ Comparison of our measurements of proper motion in right ascension
(upper panel) and declination (lower panel) with the values listed in the literature. Green dots are from \citealt{Ivanov2013}, blue dots from \citealt{Lepine2005,Lepine2008}, and
the red dot is VVV BD 001 \citep{Beamin2013}. The error bars of each object are smaller than the point size}
  \label{Fig:pmra}
\end{figure}

We obtained improved proper motion and parallaxes for
the first time for these 17 sources, also provide revised
proper motion and parallax for an additional source (VVV
BD 001).
In Fig. \ref{Fig:pmra} we show a comparison between our estimates and previous results of proper motion in RA and DEC. The
green points show stars from \citealt{Ivanov2013} and blue
dots stars from \citealt{Lepine2005,Lepine2008}, and the red dot is VVVBD-001.
The red solid line show a 1:1 relation, the black
dashed line is the best fit with blue shade marking 1-$\sigma$
interval.
We reach milli and sub-milli arcsec precision in a couple
of sources,comparable with space based observations
and results from state of the art ground based astrometric
projects.
A summary of our astrometric results is shown in Table \ref{Tab:astrometry}. 
\begin{table*}
\small
\label{Tab:astrometry}     
\centering                         
\begin{tabular}{c@{     }  c@{          } c@{       } c@{     } c@{     } c@{     } c@{     } c@{     }}
\hline\hline                 
 Name  & $\pi$ & $\sigma _\pi$ & $\mu _\alpha$cos\,$\delta$ &  $\sigma _{\mu _\alpha cos\,\delta}$&  $\mu _\delta$ &  $\sigma_{\mu _\delta}$ & K$\rm _S$  \\
           & mas &mas & mas\,yr$^{-1}$ & mas\,yr$^{-1}$ & mas\,yr$^{-1}$ & mas\,yr$^{-1}$  & mag\\
\hline     
2MASS J13124062-6408313 B & 7.66 & 2.88 & -494.80 & 1.90 & 61.10 & 1.41 & 13.297\\
2MASS J13314733-6305061 & 7.08 & 3.25 & -571.92 & 1.81 & 108.30 & 1.92 & 12.917\\
L 149-77 B & 22.20 & 2.31 & 173.25 & 1.30 & 193.46 & 1.11 & 10.418\\
LHS 2881 B & 37.87 & 2.08 & -537.41 & 1.10 & -537.68 & 1.20 & 10.863\\
2MASS J14570297-5943468 & 15.44 & 1.67 & -388.77 & 0.87 & -472.73 & 0.96 & 11.780\\
L 200-41 B & 8.53 & 2.57 & -169.60 & 1.52 & -142.38 & 1.73 & 10.280\\
2MASS J15291530-5620406 & 12.58 & 2.89 & -583.73 & 2.00 & -129.01 & 1.81 & 12.480 \\
2MASS J15495475-5524582 & 17.69 & 1.64 & -521.57 & 0.82 & -273.14 & 0.97 & 11.753\\
VVV J172640.22-273803.13 & 53.93 & 1.20 & -546.38 & 0.74 & -322.29 & 0.82 & 12.194\\
2MASS J17433729-2303586 & 9.59 & 0.76 & -86.19 & 0.60 & -477.57 & 0.75  & 11.031\\
2MASS J17442929-2044481 & 25.91 & 2.10 & -242.02 & 1.60 & -689.99 & 1.90 & 10.599\\
2MASS J17505469-2734335 & 14.55 & 1.60 & -397.35 & 1.00 & -358.95 & 1.00 & 10.947\\
2MASS J17512332-3505258 & 3.91 & 0.98 & -150.74 & 0.60 & -129.30 & 0.60 & 11.339\\
2MASS J17521105-3916546 & 18.04 & 1.22 & -88.90 & 0.82 & -367.84 & 0.89 & 10.988\\
MACHO 124.22158.2900 & 9.55 & 1.22 & -143.11 & 0.91 & -245.84 & 0.89 &11.562 \\
MACHO 124.22158.2910 & 5.91 & 1.71 & -141.50 & 1.30 & -246.39 & 1.11 & 12.09\\
J18092457-4104351 & 31.01 & 2.65 & 173.58 & 1.44 & -386.85 & 1.83 & 10.608\\
\hline
 & & $<\sigma _\pi>$  & &$ <\sigma \mu _{\alpha}>$  & &$<\sigma \mu _{\delta}>$\\
\hline
 &  &1.92& &1.2 & &1.23\\
\hline                                   
\end{tabular}
\caption{Astrometry for the eighteen sources in this study
Last row shows the average of the errors in parallax and proper motion at each coordinate.}            
\end{table*}

\subsection{Remarkable objects}
\begin{figure}[!t]
  \includegraphics[width=\columnwidth]{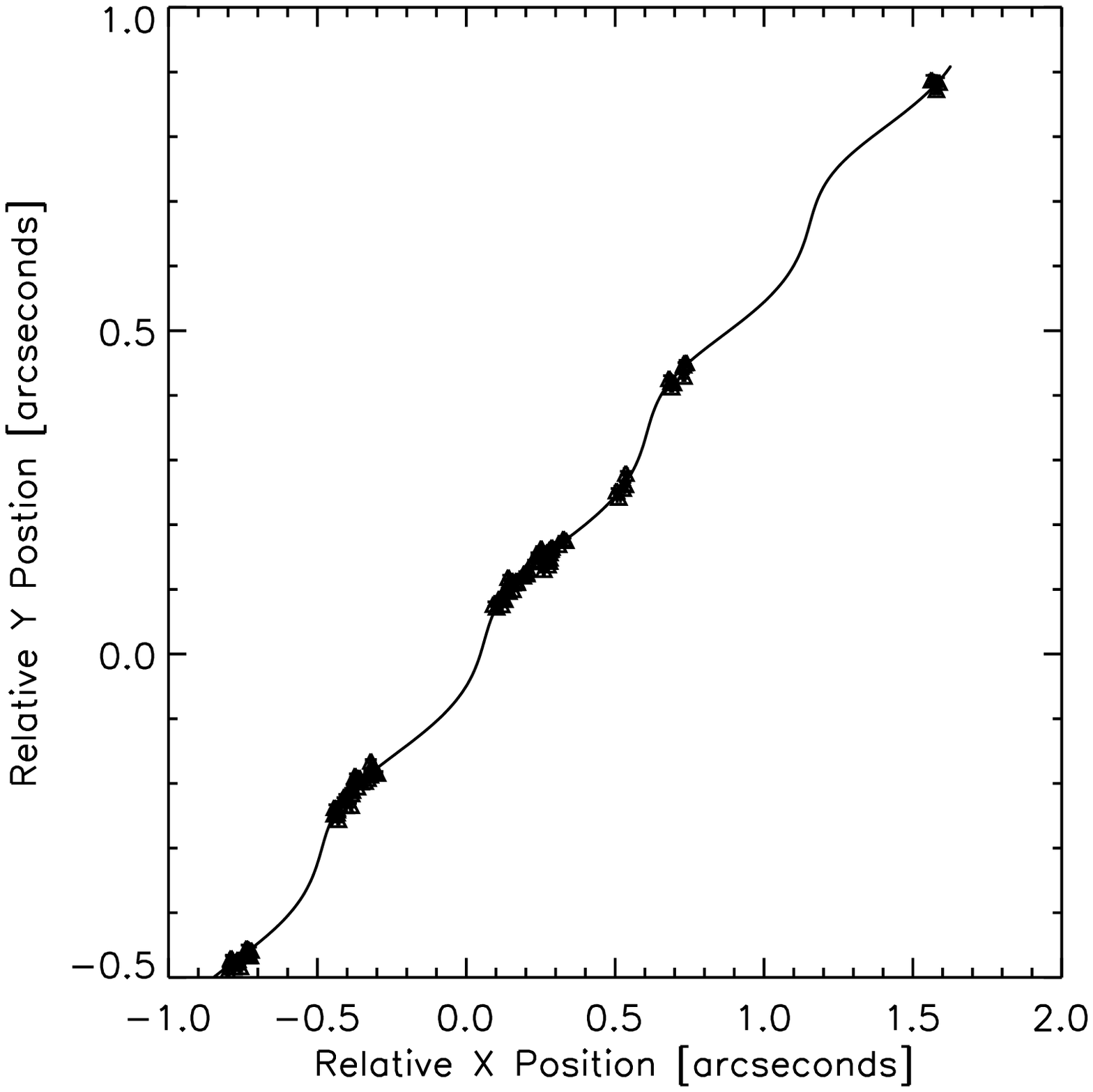}
  \includegraphics[width=\columnwidth]{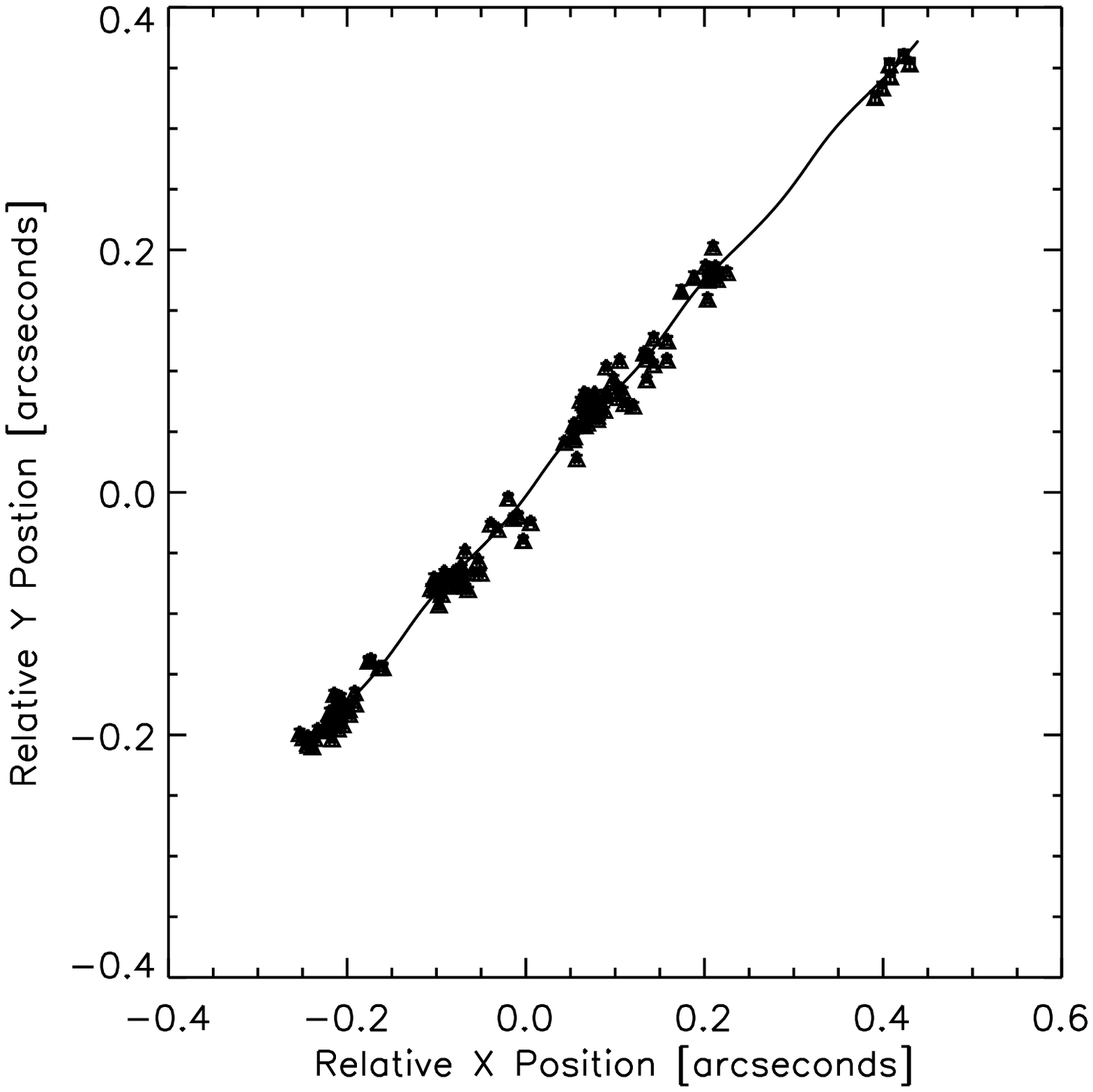}
  \caption{Motion on the sky for the closest and farthest source in this study and the best fit, including proper motion and parallax. Upper panel, VVV BD001, the estimated distance is 18.55 $\pm ^{0.42}_{0.4}$\,pc. Lower panel 2MASS J17512332-3505258, located at a distance of 255$\pm ^{86}_{51}$\,pc.}
  \label{fig:par1}
\end{figure}

2MASS J13124062-6408313 is a wide binary most likely
a member of the Halo, given the colours, and large
tangential velocity (Vtan $>$300 Km/s), there are another
five sources that have tangential velocities greater than
200 km/s making them most likely members of the
galactic halo, most of them also show unusual colours
making them good candidates for subdwarf or
unresolved binaries
We tested our parallax against the new Tycho-Gaia astrometric solution (TGAS)  \citep{TGAS}, in the catalog we found parallaxes for L 149-77 and L
200-41 and we present parallaxes for L 149-77 B, L 200-41 B. The result are in agreement within 1.5$\sigma$, 1$\sigma$ with the Gaia results for the primaries.
Examples of the source motion, including parallax for the
nearest and farthest away objects in our study are
shown in Fig \ref{fig:par1}.
Finally, 2MASS J17433729-2303586 has a small chance
to produce a microlensing event in the next 5 years, a
detailed analysis and predictions for this source are
being carried on.

\section{Conclusions}
We present improved NIR astrometry including parallaxes
for eighteen sources.
We found five sources that are more likely nearby halo
objects, in particular two wide binary system, and one
probable unresolved binary.
Spectroscopic follow up for these sources is encouraged,
particularly RV measurements for the wide binaries with
high tangential velocity.
Our results show consistent distances for sources up to
distances of $\sim$250 pc. And in the case of co-moving
companions our parallax measurements agree within the
errors with the parallax calculated for the primary source
by the Gaia mission.
Finally we found one source that might produce a
microlensing event in the upcoming years, we encourage
high spatial resolution imaging, spectroscopic and time
series photometric follow up for this target.
\smallskip
\smallskip
\it{Acknowledgments} This study was supported by programa Comit\'e Mixto ESO-Gobierno de Chile.
This publication makes use of data products from the Two Micron All Sky Survey, which is a joint project of the University of Massachusetts and the Infrared Processing and Analysis Center/California Institute of Technology, funded by the National Aeronautics and Space Administration and the National Science Foundation, and SIMBAD database, operated at CDS, Strasbourg, France.

\end{document}